\documentstyle[12pt]{article}

\begin{document}

\title{Lorentz Covariant Distributions with the Spectral Conditions}

\author{Yury M. Zinoviev\thanks{This work was supported in part by the Russian Foundation
 for Basic Research (Grant No. 05 - 01 - 00312) and Scientific Schools 672.2006.1}}

\date{}
\maketitle

\vskip 1cm

Steklov Mathematical Institute, Gubkin St. 8, 119991, Moscow, Russia,

 e - mail: zinoviev@mi.ras.ru

\vskip 1cm

\noindent {\bf Abstract.} The Lorentz covariant tempered distributions with the
supports in the product of the closed upper light cones are described.

\vskip 2cm

\section{Introduction}
\setcounter{equation}{0}

G\aa rding and Wightman \cite{1} formulated the physical views of the quantum fields
as the axioms system. The vacuum expectation values of the quantum fields products
would be the Fourier transforms of the Lorentz covariant tempered distributions with
the supports in the product of the closed upper light cones \cite{2}. The Lorentz
invariant tempered distributions of one variable are studied in the papers \cite{3},
\cite{4}. The description of the Lorentz invariant tempered distributions with the
supports for one argument only lying in the hyperboloid $\{ x \in {\bf R}^{4}: x^{0}
> 0, (x,x) \geq \mu > 0\} $ is obtained in the paper \cite{5}. The authors of the
papers \cite{3} - \cite{5} wanted to describe the Lorentz invariant distributions in
terms of the distributions given on the Lorentz group orbit space. This orbit space
has a complicated structure. It is noted \cite{6} that a tempered distribution with
a support in the closed upper light cone may be represented as the action of the
wave operator in some power on a differentiable function with a support in the
closed upper light cone. For the description of the Lorentz covariant differentiable
functions the boundary of the closed upper light cone is not important. The measure
of this boundary is zero.

In this paper we obtain the description of the Lorentz covariant tempered
distributions with the supports in the product of the closed upper light cones. Thus
we obtain the description of the vacuum expectation values of the products of the
quantum fields satisfying the properties of Lorentz covariance and spectral
condition \cite{2}.

\section{Lorentz covariance and spectral condition}

For a complex $2\times 2$ - matrix
\begin{equation}
\label{2.1}
A =  \left( \begin{array}{cc}

A_{11} & A_{12} \\

A_{21} & A_{22}

\end{array} \right).
\end{equation}
we define the following $2\times 2$ - matrices
\begin{equation}
\label{2.2}
A^{T} = \left( \begin{array}{cc}

A_{11} & A_{21} \\

A_{12} & A_{22}

\end{array} \right),
\bar{A} = \left( \begin{array}{cc}

\bar{A}_{11} & \bar{A}_{12} \\

\bar{A}_{21} & \bar{A}_{22}

\end{array} \right),
A^{\ast} = (\bar{A})^{T}.
\end{equation}
The matrix (\ref{2.1}) is said to be Hermitian if $A^{\ast} = A$. Let us choose the
basis of the Hermitian $2\times 2$ - matrices
\begin{equation}
\label{2.3}
\sigma^{0} = \left( \begin{array}{cc}

1 & 0 \\

0 & 1

\end{array} \right),
\sigma^{1} = \left( \begin{array}{cc}

0 & 1 \\

1 & 0

\end{array} \right),
\sigma^{2} = \left( \begin{array}{cc}

0 & - i \\

i &   0

\end{array} \right),
\sigma^{3} = \left( \begin{array}{cc}

1 & 0 \\

0 & - 1

\end{array} \right).
\end{equation}
We relate with a vector $x \in {\bf R}^{4}$ a Hermitian matrix
\begin{equation}
\label{2.4}
\tilde{x} = \sum_{\mu = 0}^{3} x^{\mu} \sigma^{\mu}
\end{equation}
We introduce the bilinear form
\begin{equation}
\label{2.5}
(x,y) = x^{0}y^{0} - \sum_{k = 1}^{3} x^{k}y^{k}.
\end{equation}
$(x,x)$ is the Minkowski metric.

The group $SL(2,{\bf C})$ consists of the complex $2\times 2$ - matrices (\ref{2.1})
with the determinant that equals one. The group $SU(2)$ is the maximal compact
subgroup of the group $SL(2,{\bf C})$. It consists of the matrices from the group
$SL(2,{\bf C})$ satisfying the equation $AA^{\ast} = \sigma^{0}$. Let us describe
the irreducible representations of the group $SU(2)$. We consider the half -
integers $l \in 1/2{\bf Z}_{+}$, i.e. $l = 0,1/2,1,3/2,...$. We define the
representation of the group $SU(2)$ on the space of the polynomials with degree less
than $2l$
\begin{equation}
\label{2.6}
T_{l}(A)\phi (z) = (A_{12}z + A_{22})^{2l}\phi (\frac{A_{11}z + A_{21}}{A_{12}z +
A_{22}}).
\end{equation}
We consider a half - integer $n = - l, - l + 1,...,l - 1,l$. We choose the
polynomial basis
\begin{equation}
\label{2.7}
\psi_{n} (z) = ((l - n)!(l + n)!)^{- 1/2}z^{l - n}
\end{equation}
The definitions (\ref{2.6}), (\ref{2.7}) imply
$$
T_{l}(A)\psi_{n} (z) = \sum_{m = - l}^{l} \psi_{m} (z)t_{mn}^{l}(A)
$$
where the polynomial
\begin{eqnarray}
\label{2.8}
t_{mn}^{l}(A) = ((l - m)!(l + m)!(l - n)!(l + n)!)^{1/2} \times \nonumber \\ \sum_{j
= - \infty}^{\infty} \frac{A_{11}^{l - m - j}A_{12}^{j}A_{21}^{m - n + j}A_{22}^{l +
n - j}}{\Gamma (j + 1)\Gamma (l - m - j + 1)\Gamma (m - n + j + 1)\Gamma (l + n - j
+ 1)}
\end{eqnarray}
where $\Gamma (z)$ is the gamma - function. The function $(\Gamma (z))^{- 1}$ equals
zero for $z = 0 , - 1, - 2,...$. Therefore the series (\ref{2.8}) is a polynomial.

The relation (\ref{2.6}) defines a representation of the group $SU(2)$. Thus the
polynomial (\ref{2.8}) defines a representation of the group $SU(2)$
\begin{equation}
\label{2.9}
t_{mn}^{l}(AB) = \sum_{k = - l}^{l} t_{mk}^{l}(A)t_{kn}^{l}(B).
\end{equation}
This $(2l + 1)$ - dimensional representation is irreducible (\cite{7}, Chapter III,
Section 2.3).

The relations (\ref{2.8}), (\ref{2.9}) have an analytic continuation to the matrices
from the group $SL(2,{\bf C})$. By making the change $j \rightarrow j + n - m$ of
the summation variable in the equality (\ref{2.8}) we have
\begin{equation}
\label{2.10}
t_{mn}^{l}(A) = t_{nm}^{l}(A^{T}).
\end{equation}
Due to (\cite{7}, Chapter III, Section 8.3) we have
\begin{equation}
\label{2.11}
t_{m_{1}n_{1}}^{l_{1}}(A)t_{m_{2}n_{2}}^{l_{2}}(A) = \sum_{l_{3} \in 1/2{\bf Z}_{+}}
\sum_{m_{3},n_{3} = - l_{3}}^{l_{3}}
C(l_{1},l_{2},l_{3};m_{1},m_{2},m_{3})C(l_{1},l_{2},l_{3};n_{1},n_{2},n_{3})t_{m_{3}n_{3}}^{l_{3}}(A)
\end{equation}
for a matrix $A \in SU(2)$. The Clebsch - Gordan coefficient
$C(l_{1},l_{2},l_{3};m_{1},m_{2},m_{3})$ is not zero only if $m_{3} = m_{1} + m_{2}$
and the half - integers $l_{1},l_{2},l_{3} \in 1/2{\bf Z}_{+}$ satisfy the triangle
condition: it is possible to construct a triangle with the sides of length
$l_{1},l_{2},l_{3}$ and an integer perimeter $l_{1} + l_{2} + l_{3}$. It means that
the half - integer $l_{3}$ is one of the half - integers $|l_{1} - l_{2}|,|l_{1} -
l_{2}| + 1,...,l_{1} + l_{2} - 1,l_{1} + l_{2}$. Let the half - integers
$l_{1},l_{2},l_{3} \in 1/2{\bf Z}_{+}$ satisfy the triangle condition. Let the half
- integers $m_{i} = - l_{i}, - l_{i} + 1,...,l_{i} - 1,l_{i}, i = 1,2, m_{3} = m_{1}
+ m_{2}$. Then due to (\cite{7}, Chapter III, Section 8.3)
\begin{eqnarray}
\label{2.12}
C(l_{1},l_{2},l_{3};m_{1},m_{2},m_{3}) = \nonumber \\ (- 1)^{l_{1} - l_{3} + m_{2}}
(2l_{3} + 1)^{1/2} [\frac{(l_{1} + l_{2} - l_{3})!(l_{1} + l_{3} - l_{2})!(l_{2} +
l_{3} - l_{1})!(l_{3} - m_{3})!(l_{3} + m_{3})!}{(l_{1} + l_{2} + l_{3} + 1)!(l_{1}
- m_{1})!(l_{1} + m_{1})!(l_{2} - m_{2})!(l_{2} + m_{2})!}]^{1/2}
\times \nonumber \\
\sum_{j = 0}^{l_{2} + l_{3} - l_{1}} \frac{(- 1)^{j}(l_{1} + m_{1} + j)!)(l_{2} +
l_{3} - m_{1} - j)!}{j! \Gamma (l_{3} - m_{3} - j + 1) \Gamma (l_{1} - l_{2} + m_{3}
+ j + 1)(l_{2} + l_{3} - l_{1} - j)!}. \nonumber \\
\end{eqnarray}
Let $dA$ be the normalized Haar measure on the group $SU(2)$. Due to (\cite{7},
Chapter III, Section 8.3)
\begin{eqnarray}
\label{2.13}
C(l_{1},l_{2},l_{3};m_{1},m_{2},m_{3})C(l_{1},l_{2},l_{3};n_{1},n_{2},n_{3}) =
\nonumber \\ (2l_{3} + 1)\int_{SU(2)} dA
t_{m_{1}n_{1}}^{l_{1}}(A)t_{m_{2}n_{2}}^{l_{2}}(A)\overline{t_{m_{3}n_{3}}^{l_{3}}(A)}.
\end{eqnarray}
The coefficients of the polynomial (\ref{2.8}) are real. By using the relations
(\ref{2.10}) and $A^{\ast} = A^{- 1}$ we can rewrite the equality (\ref{2.13}) as
\begin{eqnarray}
\label{2.14}
C(l_{1},l_{2},l_{3};m_{1},m_{2},m_{3})C(l_{1},l_{2},l_{3};n_{1},n_{2},n_{3}) =
\nonumber \\ (2l_{3} + 1)\int_{SU(2)} dA
t_{m_{1}n_{1}}^{l_{1}}(A)t_{m_{2}n_{2}}^{l_{2}}(A)t_{n_{3}m_{3}}^{l_{3}}(A^{- 1}).
\end{eqnarray}
If the half - integers $l_{1},l_{2},l_{3} \in 1/2{\bf Z}_{+}$ satisfy the triangle
condition, then due to (\cite{7}, Chapter III, Section 8.3) we have
\begin{equation}
\label{2.15}
C(l_{1},l_{2},l_{3};l_{1},- l_{2},l_{1} - l_{2}) = [\frac{(2l_{3} +
1)(2l_{1})!(2l_{2})!}{(l_{1} + l_{2} - l_{3})!(l_{1} + l_{2} + l_{3} + 1)!}]^{1/2}.
\end{equation}
Let us choose the half - integers $n_{1} = l_{1}, n_{2} = - l_{2}$ in the equality
(\ref{2.14}). Then the relation (\ref{2.14}) and the invariance of the Haar measure
$dA$ imply
\begin{equation}
\label{2.16}
\sum_{n_{1} = - l_{1}}^{l_{1}} \sum_{n_{2} = - l_{2}}^{l_{2}}
t_{m_{1}n_{1}}^{l_{1}}(A)t_{m_{2}n_{2}}^{l_{2}}(A)C(l_{1},l_{2},l_{3};n_{1},n_{2},m_{3})
= \sum_{n_{3} = - l_{3}}^{l_{3}}
C(l_{1},l_{2},l_{3};m_{1},m_{2},n_{3})t_{n_{3}m_{3}}^{l_{3}}(A).
\end{equation}
The relation (\ref{2.16}) has an analytic continuation to any matrix from the group
$SL(2,{\bf C})$.

For any natural numbers $m.n$ and the half - integers $l_{1},...,l_{n +
1},\dot{l}_{1},...,\dot{l}_{n + 1} \in 1/2{\bf Z}_{+}; m_{i} = - l_{i}, - l_{i} +
1,...,l_{i} - 1,l_{i}, \dot{m}_{i} = - \dot{l}_{i}, - \dot{l}_{i} +
1,...,\dot{l}_{i} - 1,\dot{l}_{i}, i = 1,...,n + 1$, we consider the set of the
tempered distributions
$$
F_{m_{1},...,m_{n + 1};\dot{m}_{1},...,\dot{m}_{n + 1}}^{l_{1},...,l_{n +
1};\dot{l}_{1},...,\dot{l}_{n + 1}}(x_{1},...,x_{m}) \in S^{\prime}({\bf R}^{4m}).
$$
This set is called a Lorentz covariant distribution if for any matrix $A \in
SL(2,{\bf C})$
\begin{eqnarray}
\label{2.17}
F_{m_{1},...,m_{n + 1};\dot{m}_{1},...,\dot{m}_{n + 1}}^{l_{1},...,l_{n +
1};\dot{l}_{1},...,\dot{l}_{n + 1}}(A\tilde{x}_{1} A^{\ast},...,A\tilde{x}_{m}
A^{\ast}) = \sum_{k_{1} = - l_{1}}^{l_{1}} \cdots \sum_{k_{n + 1} = - l_{n +
1}}^{l_{n + 1}} \sum_{\dot{k}_{1} = - \dot{l}_{1}}^{\dot{l}_{1}} \cdots
\sum_{\dot{k}_{n + 1} = - \dot{l}_{n + 1}}^{\dot{l}_{n + 1}} \nonumber \\
(\prod_{i = 1}^{n + 1}
t_{m_{i}k_{i}}^{l_{i}}(A)t_{\dot{m}_{i}\dot{k}_{i}}^{\dot{l}_{i}}(\bar{A}))F_{k_{1},...,k_{n
+ 1};\dot{k}_{1},...,\dot{k}_{n + 1}}^{l_{1},...,l_{n +
1};\dot{l}_{1},...,\dot{l}_{n + 1}}(\tilde{x}_{1},...,\tilde{x}_{m})
\end{eqnarray}
where $2\times 2$ - matrix $\tilde{x}$ is given by the relation (\ref{2.4}). The
half - integers $l_{1},...,l_{n + 1},\dot{l}_{1},...,\dot{l}_{n + 1}$ in the
relation (\ref{2.17}) are not arbitrary. Let us choose the matrix $A = - \sigma^{0}$
in the equality (\ref{2.17}). For this matrix
$$
A\tilde{x}_{j} A^{\ast} = \tilde{x}_{j}, j = 1,...,m.
$$
The definition (\ref{2.8}) implies
$$
t_{m_{j}k_{j}}^{l_{j}}(- \sigma^{0}) = (- 1)^{2l_{j}}\delta_{m_{j}k_{j}}.
$$
Hence the equality (\ref{2.17}) is valid for the matrix $A = - \sigma^{0}$ if
$$
(- 1)^{2(l_{1} + \cdots + l_{n + 1} + \dot{l}_{1} + \cdots + \dot{l}_{n + 1})} = 1
$$
i.e. the sum $l_{1} + \cdots + l_{n + 1} + \dot{l}_{1} + \cdots + \dot{l}_{n + 1}$
is an integer. This condition is supposed below.

Due to the paper \cite{6} we obtain the representation for a tempered distribution
with a support in the closed upper light cone
$$
\overline{V}_{+} = \{ x \in {\bf R}^{4}: x^{0} \geq 0, (x,x) \geq 0 \}.
$$
\noindent {\bf Lemma.} {\it Let a tempered distribution} $F(x) \in S^{\prime}({\bf
R}^{4})$ {\it have a support in the closed upper light cone. There is a natural
number} $q$ {\it such that}
\begin{equation}
\label{2.18}
F(x) = (\partial_{x}, \partial_{x})^{q} f(x),
\end{equation}
\begin{equation}
\label{2.19}
(\partial_{x}, \partial_{x}) = \left( \frac{\partial}{\partial x^{0}}\right)^{2} -
\sum_{k = 1}^{3} \left( \frac{\partial}{\partial x^{k}}\right)^{2}
\end{equation}
{\it where a differentiable function} $f(x)$ {\it with a support in the closed upper
light cone is polynomial bounded.}

\noindent {\it Proof.} Let us introduce the step function
\begin{equation}
\label{2.20} \theta (x) = \left\{ {1, \hskip 0,5cm x \geq 0,} \atop {0, \hskip 0,5cm x <
0.} \right.
\end{equation}
Due to (\cite{8}, Section 30, formulas (126), (146))
\begin{equation}
\label{2.21}
(8\pi)^{- 1} (\partial_{x}, \partial_{x})^{2} (\theta (x^{0})\theta ((x,x))) =
\delta (x).
\end{equation}
The relation (\ref{2.21}) implies for any natural number $q \geq 2$
\begin{equation}
\label{2.22}
(2\pi 4^{q - 1}(q - 2)!(q - 1)!)^{- 1} (\partial_{x}, \partial_{x})^{q} ((x,x)^{q -
2}\theta (x^{0})\theta ((x,x))) = \delta (x).
\end{equation}
For any vector $x \in \overline{V}_{+}$ the intersection of the support of the
function
$$
(x - y,x - y)^{q - 2}\theta (x^{0} - y^{0})\theta ((x - y,x - y))
$$
with respect of the variable $y$ with the cone $\overline{V}_{+}$ is compact. This
function is $2(q - 3)$ times differentiable. Let a tempered distribution $F(x) \in
S^{\prime}({\bf R}^{4})$ have a support in the cone $\overline{V}_{+}$. Then for a
sufficiently large natural number $q$ the function
\begin{equation}
\label{2.23}
f(x) = (2\pi 4^{q - 1}(q - 2)!(q - 1)!)^{- 1} \int d^{4}y F(y)(x - y,x - y)^{q -
2}\theta (x^{0} - y^{0})\theta ((x - y,x - y))
\end{equation}
is differentiable and has a support in the cone $\overline{V}_{+}$. The function
(\ref{2.23}) is polynomial bounded. The lemma is proved.

Let us consider the relation (\ref{2.17}) for the simplest case $m = n = 1$.

\noindent {\bf Proposition.} {\it Any tempered distribution}
$F_{m_{1},m_{2};\dot{m}_{1},\dot{m}_{2}}^{l_{1},l_{2};\dot{l}_{1},\dot{l}_{2}}(x)
\in S^{\prime}({\bf R}^{4})$ {\it with a support in the closed upper light cone
satisfying the covariance relation} (\ref{2.17}) {\it for} $m = n = 1$ {\it has the
following form}
\begin{eqnarray}
\label{2.24}
\int
d^{4}xF_{m_{1},m_{2};\dot{m}_{1},\dot{m}_{2}}^{l_{1},l_{2};\dot{l}_{1},\dot{l}_{2}}(x)\phi
(x) = \sum_{l_{3} \in 1/2{\bf Z}_{+}} \sum_{m_{3},\dot{m}_{3} = - l_{3}}^{l_{3}}
C(l_{1},l_{2},l_{3};m_{1},m_{2},m_{3})C(\dot{l}_{1},\dot{l}_{2},l_{3};\dot{m}_{1},\dot{m}_{2},\dot{m}_{3})
\times \nonumber \\ \int d^{4}x\theta (x^{0})\theta
((x,x))t_{m_{3}\dot{m}_{3}}^{l_{3}}(\tilde{x})
f^{l_{1},l_{2},l_{3};\dot{l}_{1},\dot{l}_{2}}((x,x)^{1/2}) (\partial_{x},
\partial_{x})^{q}\phi (x) \nonumber \\
\end{eqnarray}
{\it where a test function} $\phi (x) \in S({\bf R}^{4})$; $q$ {\it is a natural
number; the coefficient}

\noindent $C(l_{1},l_{2},l_{3};m_{1},m_{2},m_{3})$ {\it is given by the relation}
(\ref{2.12}); $2\times 2$ - {\it matrix} $\tilde{x}$ {\it is given by the relation}
(\ref{2.4}); {\it the polynomial} $t_{mn}^{l}(A)$ {\it is given by the relation}
(\ref{2.8}); {\it the differentiable function}
$f^{l_{1},l_{2},l_{3};\dot{l}_{1},\dot{l}_{2}}(s)$ {\it with a support in the
positive semi - axis is polynomial bounded.}

\noindent {\it Proof.} The coefficient $C(l_{1},l_{2},l_{3};m_{1},m_{2},m_{3})$ is
not zero when the half - integers $l_{1},l_{2},l_{3} \in 1/2{\bf Z}_{+}$ satisfy the
triangle condition. Hence the sums in the right - hand side of the equality
(\ref{2.24}) are finite.

The function $f^{l_{1},l_{2},l_{3};\dot{l}_{1},\dot{l}_{2}}(s)$ is polynomial
bounded. Hence the integrals in the right - hand side of the equality (\ref{2.24})
are absolutely convergent. The right - hand side of the equality defines the
tempered distribution from $S^{\prime}({\bf R}^{4})$.

The relations (\ref{2.9}), (\ref{2.16}) imply that the right - hand side of the
equality (\ref{2.24}) satisfies the covariance relation (\ref{2.17}) for $m = n =
1$.

Let us prove that any tempered distribution from $S^{\prime}({\bf R}^{4})$ with a
support in the cone $\overline{V}_{+}$ satisfying the covariance relation
(\ref{2.17}) for $m = n = 1$ has the form (\ref{2.24}).

If a support of a tempered distribution
$F_{m_{1},m_{2};\dot{m}_{1},\dot{m}_{2}}^{l_{1},l_{2};\dot{l}_{1},\dot{l}_{2}}(x)
\in S^{\prime}({\bf R}^{4})$ is in the cone $\overline{V}_{+}$, then the
representation (\ref{2.18}) is valid where the differentiable function
$f_{m_{1},m_{2};\dot{m}_{1},\dot{m}_{2}}^{l_{1},l_{2};\dot{l}_{1},\dot{l}_{2}}(x)$
is given by the equality (\ref{2.23}).

For the open upper light cone
$$
V_{+} = \{ x \in {\bf R}^{4}: x^{0} > 0, (x,x) > 0 \}
$$
we introduce the coordinates
\begin{equation}
\label{2.25}
\tilde{x} = \mu g(t,z)g(t,z)^{\ast}
\end{equation}
where $\mu$ is a positive number and $2\times 2$ - matrix
\begin{equation}
\label{2.26}
g(t,z) =  \left( \begin{array}{cc}

t^{- 1} & 0 \\

      z & t

\end{array} \right).
\end{equation}
$t$ is a positive number and $z$ is a complex number. The equality (\ref{2.25}) is a
consequence of Gauss decomposition for a Hermitian positive definite $2\times 2$ -
matrix.

Let us introduce the function
\begin{eqnarray}
\label{2.27}
f_{m_{1},m_{2};\dot{m}_{1},\dot{m}_{2}}^{l_{1},l_{2};\dot{l}_{1},\dot{l}_{ 2}}(\mu,
g(t,z)) = \sum_{k_{1} = - l_{1}}^{l_{1}} \sum_{k_{2} = - l_{2}}^{l_{2}}
\sum_{\dot{k}_{1} = - \dot{l}_{1}}^{\dot{l}_{1}}
\sum_{\dot{k}_{2} = - \dot{l}_{2}}^{\dot{l}_{2}} \nonumber \\
(\prod_{i = 1}^{2} t_{m_{i}k_{i}}^{l_{i}}(g(t,z)^{-
1})\overline{t_{\dot{m}_{i}\dot{k}_{i}}^{\dot{l}_{i}}(g(t,z)^{- 1})})
f_{k_{1},k_{2};\dot{k}_{1},\dot{k}_{2}}^{l_{1},l_{2};\dot{l}_{1},\dot{l}_{2 }}(\mu
g(t,z)g(t,z)^{\ast}).
\end{eqnarray}
The function
$f_{m_{1},m_{2};\dot{m}_{1},\dot{m}_{2}}^{l_{1},l_{2};\dot{l}_{1},\dot{l}_{2}}(\tilde{x})$
defined by the relation (\ref{2.23}) satisfies the covariance relation (\ref{2.17})
for $m = n = 1$. Hence for any matrix $A$ of type (\ref{2.26}) we have
\begin{equation}
\label{2.28}
f_{m_{1},m_{2};\dot{m}_{1},\dot{m}_{2}}^{l_{1},l_{2};\dot{l}_{1},\dot{l}_{ 2}}(\mu,
Ag(t,z)) =
f_{m_{1},m_{2};\dot{m}_{1},\dot{m}_{2}}^{l_{1},l_{2};\dot{l}_{1},\dot{l}_{ 2}}(\mu,
g(t,z)).
\end{equation}

Let us choose the basis of the Lie algebra of the group of the matrices (\ref{2.26})
$$
a_{1} = \left( \begin{array}{cc}

0 & 0 \\

1 & 0

\end{array} \right),
a_{2} = \left( \begin{array}{cc}

0 & 0 \\

i & 0

\end{array} \right),
a_{3} = 1/2\left( \begin{array}{cc}

1 &   0 \\

0 & - 1

\end{array} \right).
$$
The equality (\ref{2.28}) implies three relations
\begin{eqnarray}
\label{2.29}
\frac{\partial}{\partial s}
f_{m_{1},m_{2};\dot{m}_{1},\dot{m}_{2}}^{l_{1},l_{2};\dot{l}_{1},\dot{l}_{ 2}}(\mu,
\exp \{ sa_{1}\} g(t,z))|_{s = 0} = t^{- 1}\frac{\partial}{\partial \hbox{Re} z}
f_{m_{1},m_{2};\dot{m}_{1},\dot{m}_{2}}^{l_{1},l_{2};\dot{l}_{1},\dot{l}_{ 2}}(\mu,
g(t,z)) = 0,\nonumber \\ \frac{\partial}{\partial s}
f_{m_{1},m_{2};\dot{m}_{1},\dot{m}_{2}}^{l_{1},l_{2};\dot{l}_{1},\dot{l}_{ 2}}(\mu,
\exp \{ sa_{2}\} g(t,z))|_{s = 0} = t^{- 1}\frac{\partial}{\partial \hbox{Im} z}
f_{m_{1},m_{2};\dot{m}_{1},\dot{m}_{2}}^{l_{1},l_{2};\dot{l}_{1},\dot{l}_{ 2}}(\mu,
g(t,z)) = 0,\nonumber \\ \frac{\partial}{\partial s}
f_{m_{1},m_{2};\dot{m}_{1},\dot{m}_{2}}^{l_{1},l_{2};\dot{l}_{1},\dot{l}_{ 2}}(\mu,
\exp \{ sa_{3}\} g(t,z))|_{s = 0} = \nonumber \\ - 1/2(\hbox{Re}
z\frac{\partial}{\partial \hbox{Re} z} + \hbox{Im} z\frac{\partial}{\partial
\hbox{Im} z} + t\frac{\partial}{\partial t})
f_{m_{1},m_{2};\dot{m}_{1},\dot{m}_{2}}^{l_{1},l_{2};\dot{l}_{1},\dot{l}_{ 2}}(\mu,
g(t,z)) = 0.
\end{eqnarray}
Due to the relations (\ref{2.29}) the function (\ref{2.27}) is independent of the
variable $g(t,z)$. Hence the definition (\ref{2.27}) implies
\begin{equation}
\label{2.30}
f_{m_{1},m_{2};\dot{m}_{1},\dot{m}_{2}}^{l_{1},l_{2};\dot{l}_{1},\dot{l}_{ 2}}(\mu,
g(t,z)) = f_{m_{1},m_{2};\dot{m}_{1},\dot{m}_{2}}^{l_{1},l_{2};\dot{l}_{1},\dot{l}_{
2}}(\mu, \sigma^{0}) =
f_{m_{1},m_{2};\dot{m}_{1},\dot{m}_{2}}^{l_{1},l_{2};\dot{l}_{1},\dot{l}_{ 2}}(\mu
\sigma^{0}).
\end{equation}
The relations (\ref{2.9}), (\ref{2.27}) and (\ref{2.30}) imply
\begin{eqnarray}
\label{2.31}
f_{m_{1},m_{2};\dot{m}_{1},\dot{m}_{2}}^{l_{1},l_{2};\dot{l}_{1},\dot{l}_{ 2}}(\mu
g(t,z)g(t,z)^{\ast}) = \sum_{k_{1} = - l_{1}}^{l_{1}} \sum_{k_{2} = - l_{2}}^{l_{2}}
\sum_{\dot{k}_{1} = - \dot{l}_{1}}^{\dot{l}_{1}}
\sum_{\dot{k}_{2} = - \dot{l}_{2}}^{\dot{l}_{2}} \nonumber \\
(\prod_{i = 1}^{2}
t_{m_{i}k_{i}}^{l_{i}}(g(t,z))\overline{t_{\dot{m}_{i}\dot{k}_{i}}^{\dot{l}_{i}}(g(t,z))})
f_{k_{1},k_{2};\dot{k}_{1},\dot{k}_{2}}^{l_{1},l_{2};\dot{l}_{1},\dot{l}_{2 }}(\mu
\sigma^{0}).
\end{eqnarray}

We choose the natural number $q$ in the relation (\ref{2.23}) such that the function

\noindent
$f_{m_{1},m_{2};\dot{m}_{1},\dot{m}_{2}}^{l_{1},l_{2};\dot{l}_{1},\dot{l}_{2}}(\tilde{x})$
is $l_{1} + l_{2} + \dot{l}_{1} + \dot{l}_{2} + 1$ times differentiable. (The
covariance relation (\ref{2.17}) supposes that the number $l_{1} + l_{2} +
\dot{l}_{1} + \dot{l}_{2}$ is nonnegative integer.) The support of this function
lies in the cone $\overline{V}_{+}$. Therefore its first $l_{1} + l_{2} +
\dot{l}_{1} + \dot{l}_{2}$ derivatives vanish on the boundary of the cone
$\overline{V}_{+}$. Hence the function
$f_{k_{1},k_{2};\dot{k}_{1},\dot{k}_{2}}^{l_{1},l_{2};\dot{l}_{1},\dot{l}_{2 }}(\mu
\sigma^{0})$ is $l_{1} + l_{2} + \dot{l}_{1} + \dot{l}_{2} + 1$ times differentiable
and its first $l_{1} + l_{2} + \dot{l}_{1} + \dot{l}_{2}$ derivatives vanish at the
point $\mu = 0$. Due to the relation (\ref{2.23}) the function
$f_{k_{1},k_{2};\dot{k}_{1},\dot{k}_{2}}^{l_{1},l_{2};\dot{l}_{1},\dot{l}_{2 }}(\mu
\sigma^{0})$ is polynomial bounded.

By making use of the coordinate substitution (\ref{2.25}) and the relation
(\ref{2.31}) we have
\begin{eqnarray}
\label{2.32}
\int
d^{4}xf_{m_{1},m_{2};\dot{m}_{1},\dot{m}_{2}}^{l_{1},l_{2};\dot{l}_{1},\dot{l}_{
2}}(\tilde{x} )\phi (\tilde{x} ) = \sum_{k_{1} = - l_{1}}^{l_{1}} \sum_{k_{2} = -
l_{2}}^{l_{2}} \sum_{\dot{k}_{1} = - \dot{l}_{1}}^{\dot{l}_{1}} \sum_{\dot{k}_{2} =
- \dot{l}_{2}}^{\dot{l}_{2}} \int_{0}^{\infty} \mu^{3} d\mu \int_{0}^{\infty} t^{-
3}dt\int d\hbox{Re} z\int d\hbox{Im} z \nonumber \\ (\prod_{i = 1}^{2}
t_{m_{i}k_{i}}^{l_{i}}(g(t,z))\overline{t_{\dot{m}_{i}\dot{k}_{i}}^{\dot{l}_{i}}(g(t,z))})
2f_{k_{1},k_{2};\dot{k}_{1},\dot{k}_{2}}^{l_{1},l_{2};\dot{l}_{1},\dot{l}_{2 }}(\mu
\sigma^{0}) \phi (\mu g(t,z)g(t,z)^{\ast}). \nonumber \\
\end{eqnarray}
The function
$f_{m_{1},m_{2};\dot{m}_{1},\dot{m}_{2}}^{l_{1},l_{2};\dot{l}_{1},\dot{l}_{2}}(\tilde{x})$
defined by the relation (\ref{2.23}) satisfies the covariance relation (\ref{2.17})
for $m = n = 1$. We substitute in the equality (\ref{2.32}) the sequence of the
functions $\phi_{\nu} (\tilde{x}) \in S({\bf R}^{4})$ converging to the distribution
$$
2(x,x)^{3/2}\delta ({\bf x})\psi ((x,x)^{1/2})
$$
where a vector ${\bf x} = (x^{1},x^{2},x^{3})$ and a function $\psi (s) \in S({\bf
R})$. Now the covariance relation (\ref{2.17}) for a matrix $A \in SU(2)$ implies
\begin{equation}
\label{2.33}
f_{m_{1},m_{2};\dot{m}_{1},\dot{m}_{2}}^{l_{1},l_{2};\dot{l}_{1},\dot{l}_{ 2}}(\mu
\sigma^{0}) = \sum_{k_{1} = - l_{1}}^{l_{1}} \sum_{k_{2} = - l_{2}}^{l_{2}}
\sum_{\dot{k}_{1} = - \dot{l}_{1}}^{\dot{l}_{1}} \sum_{\dot{k}_{2} = -
\dot{l}_{2}}^{\dot{l}_{2}} (\prod_{i = 1}^{2}
t_{m_{i}k_{i}}^{l_{i}}(A)\overline{t_{\dot{m}_{i}\dot{k}_{i}}^{\dot{l}_{i}}(A)})
f_{k_{1},k_{2};\dot{k}_{1},\dot{k}_{2}}^{l_{1},l_{2};\dot{l}_{1},\dot{l}_{2 }}(\mu
\sigma^{0}).
\end{equation}
We integrate the relation (\ref{2.33}) with the normalized Haar measure $dA$ on the
group $SU(2)$. The relations (\ref{2.11}), (\ref{2.13}) imply
\begin{eqnarray}
\label{2.34}
f_{m_{1},m_{2};\dot{m}_{1},\dot{m}_{2}}^{l_{1},l_{2};\dot{l}_{1},\dot{l}_{ 2}}(\mu
\sigma^{0}) = \sum_{l_{3} \in 1/2{\bf Z}_{+}} \sum_{m_{3},k_{3} = - l_{3}}^{l_{3}}
\sum_{k_{1} = - l_{1}}^{l_{1}} \sum_{k_{2} = - l_{2}}^{l_{2}} \sum_{\dot{k}_{1} = -
\dot{l}_{1}}^{\dot{l}_{1}} \sum_{\dot{k}_{2} = - \dot{l}_{2}}^{\dot{l}_{2}}
C(l_{1},l_{2},l_{3};m_{1},m_{2},m_{3})\times \nonumber \\ C(\dot{l}_{1},
\dot{l}_{2}, l_{3}; \dot{m}_{1}, \dot{m}_{2}, m_{3}) (2l_{3} + 1)^{-
1}C(l_{1},l_{2},l_{3};k_{1},k_{2},k_{3})C(\dot{l}_{1}, \dot{l}_{2}, l_{3};
\dot{k}_{1}, \dot{k}_{2}, k_{3})
f_{k_{1},k_{2};\dot{k}_{1},\dot{k}_{2}}^{l_{1},l_{2};\dot{l}_{1},\dot{l}_{2 }}(\mu
\sigma^{0}). \nonumber \\
\end{eqnarray}
Let us introduce the function
\begin{eqnarray}
\label{2.35}
f^{l_{1},l_{2},l_{3};\dot{l}_{1},\dot{l}_{ 2}}(\mu) = \sum_{k_{1} = - l_{1}}^{l_{1}}
\sum_{k_{2} = - l_{2}}^{l_{2}} \sum_{\dot{k}_{1} = - \dot{l}_{1}}^{\dot{l}_{1}}
\sum_{\dot{k}_{2} = - \dot{l}_{2}}^{\dot{l}_{2}} \sum_{k_{3} = - l_{3}}^{l_{3}}
(2l_{3} + 1)^{- 1}\times \nonumber
\\ C(l_{1},l_{2},l_{3};k_{1},k_{2},k_{3})C(\dot{l}_{1}, \dot{l}_{2}, l_{3};
\dot{k}_{1}, \dot{k}_{2}, k_{3}) \mu^{- 2l_{3}}
f_{k_{1},k_{2};\dot{k}_{1},\dot{k}_{2}}^{l_{1},l_{2};\dot{l}_{1},\dot{l}_{2 }}(\mu
\sigma^{0}).
\end{eqnarray}
The function
$f_{m_{1},m_{2};\dot{m}_{1},\dot{m}_{2}}^{l_{1},l_{2};\dot{l}_{1},\dot{l}_{2}}(\mu
\sigma^{0})$ is $l_{1} + l_{2} + \dot{l}_{1} + \dot{l}_{2} + 1$ times differentiable
and its first $l_{1} + l_{2} + \dot{l}_{1} + \dot{l}_{2}$ vanish at the point $\mu =
0$. Hence the function (\ref{2.35}) is differentiable. The function
$f_{m_{1},m_{2};\dot{m}_{1},\dot{m}_{2}}^{l_{1},l_{2};\dot{l}_{1},\dot{l}_{2}}(\mu
\sigma^{0})$ is polynomial bounded. Hence the function (\ref{2.35}) is polynomial
bounded.

By making use of the relations (\ref{2.34}), (\ref{2.35}) and the relations
(\ref{2.10}), (\ref{2.16}) for a matrix (\ref{2.26}) we can rewrite the equality
(\ref{2.32}) as
\begin{eqnarray}
\label{2.36}
\int
d^{4}xf_{m_{1},m_{2};\dot{m}_{1},\dot{m}_{2}}^{l_{1},l_{2};\dot{l}_{1},\dot{l}_{
2}}(\tilde{x} )\phi (\tilde{x} ) = \sum_{l_{3} \in 1/2{\bf Z}_{+}}
\sum_{m_{3},\dot{m}_{3} = - l_{3}}^{l_{3}} \int_{0}^{\infty} \mu^{3} d\mu
\int_{0}^{\infty} t^{- 3}dt\int d\hbox{Re} z\int d\hbox{Im} z \nonumber \\
C(l_{1},l_{2},l_{3};m_{1},m_{2},m_{3}) C(\dot{l}_{1},
\dot{l}_{2}, l_{3}; \dot{m}_{1}, \dot{m}_{2}, \dot{m}_{3}) \times \nonumber \\
t_{m_{3}\dot{m}_{3}}^{l_{3}}(\mu g(t,z)g(t,z)^{\ast})
2f^{l_{1},l_{2},l_{3};\dot{l}_{1},\dot{l}_{2 }}(\mu ) \phi (\mu
g(t,z)g(t,z)^{\ast}).
\end{eqnarray}
By making use of the coordinate substitution inverse of the substitution
(\ref{2.25}) and the equality (\ref{2.18}) we obtain the equality (\ref{2.24}). The
proposition is proved.

Let us consider a tempered distribution with a support in the cone
$\overline{V}_{+}$ satisfying the covariance relation (\ref{2.17}) for $m = 1$, $n =
0$. This case corresponds to the equality (\ref{2.24}) with $l_{2} = \dot{l}_{2} =
0$, $m_{2} = \dot{m}_{2} = 0$. The relation (\ref{2.12}) implies
\begin{equation}
\label{2.37}
C(l_{1},0,l_{3};m_{1},0,m_{3}) = \delta_{l_{1}l_{3}} \delta_{m_{1}m_{3}}.
\end{equation}

Let us consider the general case.

\noindent {\bf Theorem.} {\it Any tempered distribution}
$$
F_{m_{1},...,m_{n + 1};\dot{m}_{1},...,\dot{m}_{n + 1}}^{l_{1},...,l_{n +
1};\dot{l}_{1},...,\dot{l}_{n + 1}}(x_{1},...,x_{n + 1}) \in S^{\prime}({\bf
R}^{4m})
$$
{\it with a support in the product} $\overline{V}_{+}^{\times m}$ {\it of the cones
satisfying the covariance relation} (\ref{2.17}) {\it has the following form}
\begin{eqnarray}
\label{2.38}
\int d^{4m}xF_{m_{1},...,m_{n + 1};\dot{m}_{1},...,\dot{m}_{n + 1}}^{l_{1},...,l_{n
+ 1};\dot{l}_{1},...,\dot{l}_{n + 1}}(x_{1},...,x_{m })\phi (x_{1},...,x_{m}) =
\nonumber \\ \sum_{k_{1} = - l_{1}}^{l_{1}} \cdots \sum_{k_{n + 1} = - l_{n +
1}}^{l_{n + 1}} \sum_{\dot{k}_{1} = - \dot{l}_{1}}^{\dot{l}_{1}} \cdots
\sum_{\dot{k}_{n + 1} = - \dot{l}_{n + 1}}^{\dot{l}_{n + 1}} \int_{0}^{\infty}
\mu^{4m - 1} d\mu \int_{0}^{\infty} t^{- 3}dt\int d\hbox{Re} z\int d\hbox{Im} z
\int_{SU(2)} dA \nonumber \\ \int d^{4m}p \delta (\sum_{j = 1}^{m} \tilde{p}_{j} -
\sigma^{0} ) (\prod_{i = 1}^{n + 1}
t_{m_{i}k_{i}}^{l_{i}}(g(t,z)A)\overline{t_{\dot{m}_{i}\dot{k}_{i}}^{\dot{l}_{i}}(g(t,z)A)})
\times \nonumber \\ 2f_{k_{1},...,k_{n + 1};\dot{k}_{1},...,\dot{k}_{n +
1}}^{l_{1},...,l_{n + 1};\dot{l}_{1},...,\dot{l}_{n + 1}}(\mu p_{1},...,\mu p_{m})
(\prod_{i = 1}^{m} (\partial_{x_{i}},
\partial_{x_{i}} )^{q})
\phi (x_{1},...,x_{m})|_{\tilde{x}_{i} = \mu g(t,z)A\tilde{p}_{i} (g(t,z)A)^{\ast}, i = 1,...,m} \nonumber \\
\end{eqnarray}
{\it where a test function} $\phi (x_{1},...,x_{m}) \in S({\bf R}^{4m})$; $dA$ {\it
is the normalized Haar measure on the group} $SU(2)$; $2\times 2$ - {\it matrix}
$g(t,z)$ {\it has the form} (\ref{2.26}); {\it the polynomial} $t_{mn}^{l}(A)$ {\it
is defined by the relation} (\ref{2.8}); $q$ {\it is a natural number}; {\it a
function} $f_{m_{1},...,m_{n + 1};\dot{m}_{1},...,\dot{m}_{n + 1}}^{l_{1},...,l_{n +
1};\dot{l}_{1},...,\dot{l}_{n + 1}}(p_{1},...,p_{m})$ {\it is polynomial bounded
and} $l_{1} + \cdots + l_{n + 1} + \dot{l}_{1} + \cdots + \dot{l}_{n + 1} + 1$ {\it
times differentiable} ({\it The covariance relation} (\ref{2.17}) {\it supposes that
the half - integer} $l_{1} + \cdots + l_{n + 1} + \dot{l}_{1} + \cdots + \dot{l}_{n
+ 1}$ {\it is a nonnegative integer}.); {\it a support of a function}
$f_{m_{1},...,m_{n + 1};\dot{m}_{1},...,\dot{m}_{n + 1}}^{l_{1},...,l_{n +
1};\dot{l}_{1},...,\dot{l}_{n + 1}}(p_{1},...,p_{m})$ {\it lies in the product}
$\overline{V}_{+}^{\times m}$ {\it of the cones}.

\noindent {\it Proof}. The group of the matrices (\ref{2.26}) is called the group
$Z_{-}D(2)$. For any matrix from the group $SL(2,{\bf C})$ the Gram decomposition is
valid
\begin{eqnarray}
\label{2.39}
\left( \begin{array}{cc}

A_{11} & A_{12} \\

A_{21} & A_{22}

\end{array} \right) =
\left( \begin{array}{cc}

t^{- 1} & 0 \\

z       & t

\end{array} \right)
\left( \begin{array}{cc}

\alpha        & \beta \\

- \bar{\beta} & \bar{\alpha}

\end{array} \right), \nonumber \\
t = (|A_{11}|^{2} + |A_{12}|^{2})^{- 1/2}, \nonumber \\
z = (\overline{A}_{11}A_{21} + \overline{A}_{12}A_{22})(|A_{11}|^{2} +
|A_{12}|^{2})^{- 1/2}, \nonumber \\
\alpha = A_{11}(|A_{11}|^{2} + |A_{12}|^{2})^{- 1/2}, \nonumber \\
\beta = A_{12}(|A_{11}|^{2} + |A_{12}|^{2})^{- 1/2}.
\end{eqnarray}
The first matrix in the right - hand side of the first equality (\ref{2.39}) $g(t,z)
\in Z_{-}D(2)$ and the second matrix $u \in SU(2)$. Let $du$ be the normalized Haar
measure on the group $SU(2)$. We consider the measure $d(g(t,z)u) = t^{-
3}dtd\hbox{Re} zd\hbox{Im} zdu$ on the group $SL(2,{\bf C})$. If this measure is
invariant under the left shifts on the group $SL(2,{\bf C})$, then the right - hand
side of the equality (\ref{2.38}) satisfies the covariance relation (\ref{2.17}) due
to the relation (\ref{2.9}). It is easy to verify that the measure $dg(t,z) = t^{-
3}dtd\hbox{Re} zd\hbox{Im} z$ is invariant under the left shifts on the group
$Z_{-}D(2)$. Hence the measure $d(g(t,z)u)$ is invariant under the left
multiplication of the matrix (\ref{2.39}) by any matrix from the group $Z_{-}D(2)$.
Due to the Gram decomposition it is sufficient to prove the invariance of the
measure $d(g(t,z)u)$ under the left multiplication of the matrix (\ref{2.39}) by any
matrix from the group $SU(2)$. Let the complex numbers $\alpha, \beta$ satisfy the
relation $|\alpha|^{2} + |\beta|^{2} = 1$. The Gram decomposition (\ref{2.39})
implies
\begin{eqnarray}
\label{2.40}
\left( \begin{array}{cc}

\alpha        & \beta \\

- \bar{\beta} & \bar{\alpha}

\end{array} \right)
\left( \begin{array}{cc}

t^{- 1} & 0 \\

z       & t

\end{array} \right)
 = \left( \begin{array}{cc}

t_{1}^{- 1} & 0 \\

z_{1}       & t_{1}

\end{array} \right)
\left( \begin{array}{cc}

\alpha_{1}        & \beta_{1} \\

- \bar{\beta}_{1} & \bar{\alpha}_{1}

\end{array} \right), \nonumber \\
t_{1} = (|\alpha t^{- 1} + \beta z|^{2} + |\beta|^{2}t^{2})^{- 1/2}, \nonumber \\
z_{1} = ((\bar{\alpha} t^{- 1} + \bar{\beta} \bar{z})(- \bar{\beta} t^{- 1} +
\bar{\alpha} z) + \bar{\alpha} \bar{\beta} t^{2})(|\alpha t^{- 1} + \beta z|^{2} +
|\beta|^{2}t^{2})^{- 1/2}, \nonumber \\
\alpha_{1} = (\alpha t^{- 1} + \beta z)(|\alpha t^{- 1} + \beta z|^{2} +
|\beta|^{2}t^{2})^{- 1/2}, \nonumber \\
\beta_{1} = \beta t(|\alpha t^{- 1} + \beta z|^{2} + |\beta|^{2}t^{2})^{- 1/2}.
\end{eqnarray}
The Haar measure $du$ is invariant under the shifts on the group $SU(2)$. In order
to prove the invariance of the measure $d(g(t,z)u)$ under the left multiplication of
the matrix (\ref{2.39}) by any matrix from the group $SU(2)$ it is sufficient to
prove the equality
\begin{equation}
\label{2.41}
t_{1}^{- 3}dt_{1}d\hbox{Re} z_{1}d\hbox{Im} z_{1}  = t^{- 3}dtd\hbox{Re} zd\hbox{Im}
z
\end{equation}
where the numbers $t_{1}$, $z_{1}$ are defined by the second and the third relations
(\ref{2.40}). We introduce the coordinates
\begin{eqnarray}
\label{2.42}
(|{\bf x}|^{2} + 1)^{1/2}\sigma^{0} + \sum_{k = 1}^{3} x^{k}\sigma^{k} =
g(t,z)g(t,z)^{\ast}, \nonumber \\
|{\bf x}|^{2} = \sum_{k = 1}^{3} (x^{k})^{2}.
\end{eqnarray}
It is easy to calculate
\begin{equation}
\label{2.43}
\frac{1}{2} (|{\bf x}|^{2} + 1)^{- 1/2}dx^{1}dx^{2}dx^{3} = t^{- 3}dtd\hbox{Re}
zd\hbox{Im} z.
\end{equation}
The mapping $g(t,z) \rightarrow g(t_{1},z_{1})$ corresponds with the rotation
$$
\left( \begin{array}{cc}

\alpha        & \beta \\

- \bar{\beta} & \bar{\alpha}

\end{array} \right) ((|{\bf x}|^{2} + 1)^{1/2}\sigma^{0} + \sum_{k = 1}^{3} x^{k}\sigma^{k})
\left( \begin{array}{cc}

\alpha        & \beta \\

- \bar{\beta} & \bar{\alpha}

\end{array} \right)^{\ast}
$$
of the vector ${\bf x} \in {\bf R}^{3}$ given by the relation (\ref{2.42}). The
invariance of the measure (\ref{2.43}) under the rotations implies the equality
(\ref{2.41}). Hence the measure $d(g(t,z)u) = t^{- 3}dtd\hbox{Re} zd\hbox{Im} zdu$
is invariant under the left shifts on the group $SL(2,{\bf C})$. Therefore the
distribution (\ref{2.38}) satisfies the covariance relation (\ref{2.17}).

Let us prove the absolute convergence of the integral (\ref{2.38}). The coefficients
of the polynomial (\ref{2.8}) are real. The relations (\ref{2.9}), (\ref{2.10})
imply
\begin{equation}
\label{2.44}
\sum_{k = - l}^{l} |t_{mk}^{l}(A)|^{2} = t_{mm}^{l}(AA^{\ast}).
\end{equation}
The Cauchy inequality and the relation (\ref{2.44}) imply
\begin{eqnarray}
\label{2.45}
|\sum_{k_{1} = - l_{1}}^{l_{1}} \cdots \sum_{k_{n + 1} = - l_{n + 1}}^{l_{n + 1}}
\sum_{\dot{k}_{1} = - \dot{l}_{1}}^{\dot{l}_{1}} \cdots \sum_{\dot{k}_{n + 1} = -
\dot{l}_{n + 1}}^{\dot{l}_{n + 1}} (\prod_{i = 1}^{n + 1}
t_{m_{i}k_{i}}^{l_{i}}(g(t,z)u)\overline{t_{\dot{m}_{i}\dot{k}_{i}}^{\dot{l}_{i}}(g(t,z)u)})
\times \nonumber \\ f_{k_{1},...,k_{n + 1};\dot{k}_{1},...,\dot{k}_{n +
1}}^{l_{1},...,l_{n +
1};\dot{l}_{1},...,\dot{l}_{n + 1}}(\mu p_{1},...,\mu p_{m})| \leq \nonumber \\
| \sum_{k_{1} = - l_{1}}^{l_{1}} \cdots \sum_{k_{n + 1} = - l_{n + 1}}^{l_{n + 1}}
\sum_{\dot{k}_{1} = - \dot{l}_{1}}^{\dot{l}_{1}} \cdots \sum_{\dot{k}_{n + 1} = -
\dot{l}_{n + 1}}^{\dot{l}_{n + 1}} \mu^{- 2\sum_{i = 1}^{n + 1} ( l_{i} +
\dot{l}_{i} )} \times \nonumber \\ |f_{k_{1},...,k_{n +
1};\dot{k}_{1},...,\dot{k}_{n + 1}}^{l_{1},...,l_{n + 1};\dot{l}_{1},...,\dot{l}_{n
+ 1}}(\mu p_{1},...,\mu p_{m})|^{2}|^{1/2} \times \nonumber \\
\left( \prod_{i = 1}^{n + 1} t_{m_{i}m_{i}}^{l_{i}}(\mu
g(t,z)g(t,z)^{\ast})t_{\dot{m}_{i} \dot{m}_{i}}^{\dot{l}_{i}} (\mu
g(t,z)g(t,z)^{\ast}) \right)^{1/2}.
\end{eqnarray}
Let us introduce the coordinates
\begin{eqnarray}
\label{2.46}
\tilde{x}_{j} = \mu g(t,z)\tilde{p}_{j} g(t,z)^{\ast}, j = 1,...,m, \nonumber \\
\tilde{p}_{m} = \sigma^{0} - \sum_{j = 1}^{m - 1} \tilde{p}_{j}, m > 1, \nonumber \\
\tilde{p}_{m} = \sigma^{0}, m = 1.
\end{eqnarray}
By summing up the equalities (\ref{2.46}) we obtain the decomposition (\ref{2.25})
for the matrix $\tilde{x}_{1} + \cdots + \tilde{x}_{m}$. The function
$f_{m_{1},...,m_{n + 1};\dot{m}_{1},...,\dot{m}_{n + 1}}^{l_{1},...,l_{n +
1};\dot{l}_{1},...,\dot{l}_{n + 1}}(p_{1},...,p_{m})$ is polynomial bounded. For the
set
$$
\{ p_{1},...,p_{m} \in \overline{V}_{+} : \sum_{j = 1}^{m} \tilde{p}_{j} =
\sigma^{0} \}
$$
the following estimate is valid
\begin{equation}
\label{2.47}
|f_{m_{1},...,m_{n + 1};\dot{m}_{1},...,\dot{m}_{n + 1}}^{l_{1},...,l_{n +
1};\dot{l}_{1},...,\dot{l}_{n + 1}}(\mu p_{1},...,\mu p_{m})| \leq C(1 + \sum_{\nu =
0}^{3} \sum_{i = 1}^{m} (\mu p_{i}^{\nu })^{2})^{N} \leq C(1 + 4m\mu^{2} )^{N}
\end{equation}
where the constant $C$ does not depend on the variables $p_{1},...,p_{m}$. The
function

\noindent $f_{m_{1},...,m_{n + 1};\dot{m}_{1},...,\dot{m}_{n + 1}}^{l_{1},...,l_{n +
1};\dot{l}_{1},...,\dot{l}_{n + 1}}(p_{1},...,p_{m})$ is differentiable $l_{1} +
\cdots + l_{n + 1} + \dot{l}_{1} + \cdots + \dot{l}_{n + 1} + 1$ times and its
support lies in the product $\overline{V}_{+}^{\times m}$ of the cones. Hence the
function

\noindent $f_{m_{1},...,m_{n + 1};\dot{m}_{1},...,\dot{m}_{n + 1}}^{l_{1},...,l_{n +
1};\dot{l}_{1},...,\dot{l}_{n + 1}}(\mu p_{1},...,\mu p_{m})$ is differentiable with
respect to the variable $\mu$ $l_{1} + \cdots + l_{n + 1} + \dot{l}_{1} + \cdots +
\dot{l}_{n + 1} + 1$ times and its first $l_{1} + \cdots + l_{n + 1} + \dot{l}_{1} +
\cdots + \dot{l}_{n + 1}$ derivatives vanish at the point $\mu = 0$. Now the
inequalities (\ref{2.45}), (\ref{2.47}) imply the absolute convergence of the
integral (\ref{2.38}). The integral (\ref{2.38}) defines the tempered distribution
from $S^{\prime}({\bf R}^{4m})$.

Let us prove the equality (\ref{2.38}). If a support of a tempered distribution
$$
F_{m_{1},...,m_{n + 1};\dot{m}_{1},...,\dot{m}_{n + 1}}^{l_{1},...,l_{n +
1};\dot{l}_{1},...,\dot{l}_{n + 1}}(x_{1},...,x_{n + 1}) \in S^{\prime}({\bf
R}^{4m})
$$
lies in the product $\overline{V}_{+}^{\times m}$ of the cones, then the relations
analogous with the relations (\ref{2.18}), (\ref{2.23}) are valid
\begin{equation}
\label{2.48}
F_{m_{1},...,m_{n + 1};\dot{m}_{1},...,\dot{m}_{n + 1}}^{l_{1},...,l_{n +
1};\dot{l}_{1},...,\dot{l}_{n + 1}}(x_{1},...,x_{m}) = (\prod_{i = 1}^{m}
(\partial_{x_{i}} ,\partial_{x_{i}})^{q}) f_{m_{1},...,m_{n +
1};\dot{m}_{1},...,\dot{m}_{n + 1}}^{l_{1},...,l_{n + 1};\dot{l}_{1},...,\dot{l}_{n
+ 1}}(x_{1},...,x_{m}),
\end{equation}
\begin{eqnarray}
\label{2.49}
f_{m_{1},...,m_{n + 1};\dot{m}_{1},...,\dot{m}_{n + 1}}^{l_{1},...,l_{n +
1};\dot{l}_{1},...,\dot{l}_{n + 1}}(x_{1},...,x_{m}) = \nonumber \\ (2\pi 4^{q -
1}(q - 2)!(q - 1)!)^{- m}  \int d^{4m}y F_{m_{1},...,m_{n +
1};\dot{m}_{1},...,\dot{m}_{n + 1}}^{l_{1},...,l_{n + 1};\dot{l}_{1},...,\dot{l}_{n
+ 1}}(y_{1},...,y_{m}) \times \nonumber \\ \prod _{i = 1}^{m} (x_{i} - y_{i},x_{i} -
y_{i})^{q - 2}\theta (x_{i}^{0} - y_{i}^{0})\theta ((x_{i} - y_{i},x_{i} - y_{i})).
\end{eqnarray}
We can choose the natural number $q$ in the relations (\ref{2.48}), (\ref{2.49})
such that the function (\ref{2.49}) is differentiable $l_{1} + \cdots + l_{n + 1} +
\dot{l}_{1} + \cdots + \dot{l}_{n + 1} + 1$ times. A support of the function
(\ref{2.49}) lies in the product $\overline{V}_{+}^{\times m}$ of the cones. The
function (\ref{2.49}) is polynomial bounded. If the tempered distribution
(\ref{2.48}) satisfies the covariance relation (\ref{2.17}), then the function
(\ref{2.49}) satisfies the covariance relation (\ref{2.17}) also. By making use of
the proof of Proposition we obtain
\begin{eqnarray}
\label{2.50}
f_{m_{1},...,m_{n + 1};\dot{m}_{1},...,\dot{m}_{n + 1}}^{l_{1},...,l_{n +
1};\dot{l}_{1},...,\dot{l}_{n + 1}}(\mu g(t,z)\tilde{p}_{1} g(t,z)^{\ast},...,\mu
g(t,z)\tilde{p}_{m} g(t,z)^{\ast}) = \nonumber \\
\sum_{k_{1} = - l_{1}}^{l_{1}} \cdots \sum_{k_{n + 1} = - l_{n + 1}}^{l_{n + 1}}
\sum_{\dot{k}_{1} = - \dot{l}_{1}}^{\dot{l}_{1}} \cdots \sum_{\dot{k}_{n + 1} = -
\dot{l}_{n + 1}}^{\dot{l}_{n + 1}} \nonumber \\ (\prod_{i = 1}^{n + 1}
t_{m_{i}k_{i}}^{l_{i}}(g(t,z))\overline{t_{\dot{m}_{i}\dot{k}_{i}}^{\dot{l}_{i}}(g(t,z))})
f_{k_{1},...,k_{n + 1};\dot{k}_{1},...,\dot{k}_{n + 1}}^{l_{1},...,l_{n +
1};\dot{l}_{1},...,\dot{l}_{n + 1}}(\mu \tilde{p}_{1},...,\mu \tilde{p}_{m}), \nonumber \\
\tilde{p}_{m} = \sigma^{0} - \sum_{j = 1}^{m - 1} \tilde{p}_{j}, m > 1, \nonumber \\
\tilde{p}_{m} = \sigma^{0}, m = 1.
\end{eqnarray}
By the definition (\ref{2.49}) the function $f_{k_{1},...,k_{n +
1};\dot{k}_{1},...,\dot{k}_{n + 1}}^{l_{1},...,l_{n + 1};\dot{l}_{1},...,\dot{l}_{n
+ 1}}(\mu p_{1},...,\mu p_{m})$ is differentiable $l_{1} + \cdots + l_{n + 1} +
\dot{l}_{1} + \cdots + \dot{l}_{n + 1} + 1$ times with respect to the variable $\mu$
and its first $l_{1} + \cdots + l_{n + 1} + \dot{l}_{1} + \cdots + \dot{l}_{n + 1}$
derivatives vanish at the point $\mu = 0$. The estimate (\ref{2.47}) is valid. By
using the relations (\ref{2.43}), (\ref{2.50}) and the coordinates (\ref{2.46}) we
have
\begin{eqnarray}
\label{2.51}
\int d^{4m}xf_{m_{1},...,m_{n + 1};\dot{m}_{1},...,\dot{m}_{n + 1}}^{l_{1},...,l_{n
+ 1};\dot{l}_{1},...,\dot{l}_{n + 1}}(x_{1},...,x_{n + 1})\phi
(\tilde{x}_{1},...,\tilde{x}_{m}) = \nonumber \\ \sum_{k_{1} = - l_{1}}^{l_{1}}
\cdots \sum_{k_{n + 1} = - l_{n + 1}}^{l_{n + 1}} \sum_{\dot{k}_{1} = -
\dot{l}_{1}}^{\dot{l}_{1}} \cdots \sum_{\dot{k}_{n + 1} = - \dot{l}_{n +
1}}^{\dot{l}_{n + 1}} \int_{0}^{\infty} \mu^{4m - 1} d\mu \int_{0}^{\infty} t^{-
3}dt\int d\hbox{Re} z\int d\hbox{Im} z \nonumber \\ \int d^{4m}p \delta (\sum_{j =
1}^{m} \tilde{p}_{j} - \sigma^{0} ) (\prod_{i = 1}^{n + 1}
t_{m_{i}k_{i}}^{l_{i}}(g(t,z))\overline{t_{\dot{m}_{i}\dot{k}_{i}}^{\dot{l}_{i}}(g(t,z))})
\times \nonumber \\ 2f_{k_{1},...,k_{n + 1};\dot{k}_{1},...,\dot{k}_{n +
1}}^{l_{1},...,l_{n + 1};\dot{l}_{1},...,\dot{l}_{n + 1}}(\mu p_{1},...,\mu p_{m})
\phi (\mu g(t,z)\tilde{p}_{1} g(t,z)^{\ast},...,\mu g(t,z)\tilde{p}_{m}
g(t,z)^{\ast}).
\end{eqnarray}
Let us insert in the equality (\ref{2.51}) a sequence of the functions $\phi_{\nu}
(x_{1},...,x_{m}) \in S({\bf R}^{4m})$ convergent to the distribution
$$
2\left( \sum_{i = 1}^{m} x_{i},\sum_{i = 1}^{m} x_{i}\right)^{3/2} \delta \left(
\sum_{i = 1}^{m} {\bf x}_{i}\right) \psi \left( \left( \sum_{i = 1}^{m}
x_{i},\sum_{i = 1}^{m} x_{i}\right)^{1/2}, x_{1},...,x_{m - 1}\right)
$$
where a function $\psi (\mu, x_{1},...,x_{m - 1}) \in S({\bf R}\times {\bf R}^{4(m -
1)})$. Now by using the covariance relation (\ref{2.17}) for the distribution
(\ref{2.51}) and a matrix $u \in SU(2)$ we have
\begin{eqnarray}
\label{2.52}
f_{m_{1},...,m_{n + 1};\dot{m}_{1},...,\dot{m}_{n + 1}}^{l_{1},...,l_{n +
1};\dot{l}_{1},...,\dot{l}_{n + 1}}(\mu \tilde{p}_{1} ,...,\mu \tilde{p}_{m}) = \nonumber \\
\sum_{k_{1} = - l_{1}}^{l_{1}} \cdots \sum_{k_{n + 1} = - l_{n + 1}}^{l_{n + 1}}
\sum_{\dot{k}_{1} = - \dot{l}_{1}}^{\dot{l}_{1}} \cdots \sum_{\dot{k}_{n + 1} = -
\dot{l}_{n + 1}}^{\dot{l}_{n + 1}} \nonumber \\ (\prod_{i = 1}^{n + 1}
t_{m_{i}k_{i}}^{l_{i}}(u)\overline{t_{\dot{m}_{i}\dot{k}_{i}}^{\dot{l}_{i}}(u)})
f_{k_{1},...,k_{n + 1};\dot{k}_{1},...,\dot{k}_{n + 1}}^{l_{1},...,l_{n +
1};\dot{l}_{1},...,\dot{l}_{n + 1}}(\mu u^{- 1}\tilde{p}_{1} u,...,\mu u^{- 1}\tilde{p}_{m} u), \nonumber \\
\tilde{p}_{m} = \sigma^{0} - \sum_{j = 1}^{m - 1} \tilde{p}_{j}, m > 1, \nonumber \\
\tilde{p}_{m} = \sigma^{0}, m = 1.
\end{eqnarray}
We integrate the equality (\ref{2.52}) with the normalized Haar measure on the group
$SU(2)$ and insert the derived equality in the equality (\ref{2.51}). Let us change
the integration variables
$$
\tilde{p}_{j} \rightarrow u\tilde{p}_{j} u^{\ast}, j = 1,..., m - 1, m > 1,
$$
$$
u\tilde{p}_{m} u^{\ast} = \tilde{p}_{m} = \sigma^{0}, m = 1.
$$
The derived equality and the relation (\ref{2.48}) yield the equality (\ref{2.38}).
The theorem is proved.

By using the relations (\ref{2.11}), (\ref{2.13}) it is possible to calculate the
integral over the group $SU(2)$ in the right - hand side of the equality
(\ref{2.38}) for $m = n = 1$. The derived equality may be rewritten in the form
(\ref{2.24}).

\end{document}